\documentclass[pra,aps,epsfig,psfig,multicols,showpacs,tightenlines]{revtex4}

\usepackage{graphics,bm}
\usepackage{graphicx}
\usepackage{amsmath, amssymb, graphics}

\newcommand{\beq}{\begin{equation}}
\newcommand{\eeq}{\end{equation}}
\newcommand{\bqa}{\begin{eqnarray}}
\newcommand{\eqa}{\end{eqnarray}}
\def\gsim{\mathrel {\vcenter {\baselineskip 0pt \kern 0pt
\hbox{$>$} \kern 0pt \hbox{$\sim$} }}}
\def\lsim{\mathrel {\vcenter {\baselineskip 0pt \kern 0pt
\hbox{$<$} \kern 0pt \hbox{$\sim$} }}}

\protect

\begin{document}
\title{A comparative analysis of Painlev$\rm\acute{e}$, Lax Pair, and
Similarity Transformation methods in obtaining the integrability conditions of
nonlinear Schr$\rm\ddot o$dinger equations}
\author{U. Al Khawaja}
\affiliation{ \it Physics Department, United Arab Emirates University, P.O. Box
17551, Al-Ain, United Arab Emirates.}

\date{\today}

\begin{abstract}
We derive the integrability conditions of nonautonomous nonlinear
Schr$\rm\ddot o$dinger equations using the Lax Pair and Similarity
Transformation methods. We present a comparative analysis of these
integrability conditions with those of the Painlev$\rm\acute{e}$
method. We show that while the Painlev$\rm\acute{e}$ integrability
conditions restrict the dispersion, nonlinearity, and
dissipation/gain coefficients to be space-independent and the
external potential to be only a quadratic function of position, the
Lax Pair and the Similarity Transformation methods allow for
space-dependent coefficients and an external potential that is not
restricted to the quadratic form. The integrability conditions of
the Painlev$\rm\acute{e}$ method are retrieved as a special case of
our general integrability conditions. We also derive the
integrability conditions of nonautonomous nonlinear Schr$\rm\ddot
o$dinger equations for two- and three-spacial dimensions.
\end{abstract}


\maketitle

\section{Introduction}
\label{intro_sec}

The question of integrability of nonautonomous nonlinear
Schr$\rm\ddot o$dinger (NLS) equations has recently been extensively
addressed due to their applications in the fields of trapped
Bose-Einstein condensates and optical solitons in fibers
\cite{serk}. Among many works that investigate the integrability of
the celebrated NLS equation
\cite{serk,dar1,dar2,liang,sim1,sim2,sim3,pan1}, He {\it et al.}
\cite{he} address the most general form of this equation, namely
\begin{equation}
f(x,t)\, \Psi_{xx}(x,t)+g(x,t)\, |\Psi (x,t)|^2 {\Psi}(x,t)+v(x,t)\,
\Psi (x,t)+i \,\gamma (x,t)\, \Psi (x,t) +i\, \Psi_{t}(x,t)=0
\label{gp},
\end{equation}
where the dispersion coefficient $f(x,t)$, nonlinearity coefficient
$g(x,t)$, gain/loss coefficient $\gamma(x,t)$, and external
potential $v(x,t)$ are real functions. Applying the
Painlev$\rm\acute{e}$ test, the authors were able to derive the
following integrability conditions
\begin{equation}
f(x,t)=f(t),\hspace{1cm}g(x,t)=g(t),\hspace{1cm}\gamma(x,t)=\gamma(t)\label{comp0},
\end{equation}
\begin{equation}
v(x,t)=v_0(t)+v_1(t)\,x+v_2(t)\,x^2\label{v2comp},
\end{equation}
where $v_0(t)$ and $v_1(t)$ are arbitrary and $v_2(t)$ is given by
\begin{equation}
4 f^3 g^2\,v_2+{ f g ({\dot f} {\dot g}+f {\ddot g})+g^2 ({\dot
f}^2-f {\ddot f})-2 f^2 {\dot g}^2}=0\label{compt}.
\end{equation}
Here and throughout, the dot represents time derivative and the subscripts
represent partial derivatives. Clearly, this shows that, according to the
Painlev$\rm\acute{e}$ test, Eq.~(\ref{gp}) is integrable only for
time-dependent dispersion, nonlinearity, gain/loss coefficients and for
quadratic external potential.

It is well-known, however, that an equation can be integrable in one sense and
nonintegrable in another sense. Therefore, we use in this paper two different
methods to investigate the integrability of Eq.~(\ref{gp}). We use first the
Lax Pair method and then the Similarity Transformation method to show that
Eq.~(\ref{gp}) is indeed integrable for more general time- and space-dependent
coefficients. We derive an integrability condition that reproduces the
conditions (\ref{comp0})-(\ref{compt}) as a special case. It is found that,
while the three methods generate different integrability conditions, when the
coefficients of Eq.~(\ref{gp}) are space- and time-dependent, the three methods
generate the same integrability condition when the coefficients are only
time-dependent.

While it is not the aim of this paper to discuss and derive exact
solutions of specific examples of integrable NLS equations, we
present in section~\ref{lax_sec} some examples corresponding to
known integrable cases. In addition, we present in
sections~\ref{lax_sec} and \ref{simi_sec}, as a counter example to
the Painlev$\rm\acute{e}$ result, two cases of NLS equations that
are indeed integrable with space- and time-dependent coefficients.

Finally, we use the Lax Pair method, in section~\ref{2d-3d}, to
generalize the integrability conditions for NLS equations of higher
spacial dimensions. To the best of our knowledge, the integrability
conditions derived here (Eqs.~(\ref{vx5}) and (\ref{vx523d})) are
presented in the literature for the first time.

\section{Lax Pair Method}
\label{lax_sec}

In the Lax Pair method for solving nonlinear partial differential
equations, a pair of $2\times2$ matrices, $\bf{U}$ and $\bf{V}$, is
first constructed (or derived). The Lax Pair is used to define a
linear system of equations for an auxiliary field $\bf \Phi$, namely
${\bf \Phi}_x={\bf U\cdot\Phi}$ and ${\bf \Phi}_t={\bf V\cdot\Phi}$.
The compatibility condition of this system, ${\bf U}_t-{\bf
V}_x+[{\bf U},{\bf V}]=0$, is required to be identical to the
nonlinear differential equation. The Darboux
transformation~\cite{salle} is then applied on the linear system to
obtain new solutions from known solutions.

We consider the NLS equation to be nonintegrable if the Lax Pair does not
exist. Our method of Lax Pair search \cite{usama_lax} provides an answer to the
question of existence of Lax Pair. In this method, we expand the Lax Pair $\bf
U$ and $\bf V$ in powers of $\Psi$ and its derivatives with unknown function
coefficients and require the compatibility condition to be equivalent to the
nonlinear differential equation. This results in a set of coupled differential
equations for the unknown coefficients. Solving this system of equations
determines the Lax Pair and results in an integrability condition on the
functions $f(x,t)$, $g(x,t)$, $\gamma(x,t)$, and $v(x,t)$. The details of this
calculation are relegated to Appendix~\ref{appendix_a}. The integrability
conditions of Eq.~(\ref{gp}) turn out to be
\begin{equation}
f(x,t)={c_1(t)\over g(x,t)^2}\label{fg},
\end{equation}
\begin{equation}
\gamma(x,t)=\frac{g_t(x,t)}{g(x,t)}-{1\over2} \frac{ {\dot c_2(t)}}{
c_2(t)} \label{gam},
\end{equation}
and
\begin{eqnarray}
&{}&f g^3 \left(f_t \left(g_t-2 g \gamma \right)-f_{tt}
g\right)+f_t^2
   g^4+2 f^3 g^3 \left(g v_{xx}-g_{x} v_{x}\right)\nonumber\\&+&f^2 g^2 \left(g \left(4
   g_t \gamma +g_{tt}\right)-2 g_t^2-2 g^2 \left(\gamma _t+2 \gamma
   ^2\right)\right)\nonumber\\&+&f^4 \left(36 g_{x}^4-48 g g_{xx} g_{x}^2+10 g^2 g_{xxx}
   g_{x}+g^2 \left(6 g_{xx}^2-g g^{(4)}\right)\right)=0.
   \label{vx5}
\end{eqnarray}
This condition shows that $f(x,t)$, $\gamma(x,t)$ and $v(x,t)$ are
determined by $g(x,t)$ and the arbitrary functions $c_1(t)$ and
$c_2(t)$. To get specific forms of integrable NLS equations, one
starts with a certain form of $g(x,t)$, $c_1(t)$, and $c_2(t)$ from
which the functions $f(x,t)$, $\gamma(x,t)$, and $v(x,t)$ will be
determined according to Eqs.~(\ref{fg}-\ref{vx5}). With such a
combination of coefficients, Eq.~(\ref{gp}) will be integrable.
Equations~(\ref{fg}-\ref{vx5}) represent a main result of the
present paper. All previously-known special cases can be derived
from these equations.
We list here some of these special cases. \\\\
{\bf Special case I: Constant and linear external potential}\\ With
the choices: $g(x,t)=1$ and $c_1(t)=c_2(t)=1$, Eq.~(\ref{gp}) takes
the form
\begin{equation}
\Psi_{xx}+|\Psi |^2 {\Psi}+\left({\over}c_3(t)+c_4(t)\,x\right)\,
\Psi
 +i\, \Psi_{t}=0 \label{gps1},
\end{equation}
where $c_3(t)$ and $c_4(t)$ are arbitrary real functions arising
from integrating Eq.~(\ref{vx5}) with respect to $x$. Clearly, with
$c_3(t)=c_4(t)=0$, the well-known homogeneous Gross-Pitaevskii
equation (GPE) is obtained. It is also established that
Eq.~(\ref{gps1}) is integrable and exact
solutions have been derived \cite{liu}.\\\\
{\bf Special case II: Harmonic potential and gain/damping term}\\
For $g(x,t)=1$, $c_1(t)=1$, and $c_2(t)=e^{\alpha\,t}$, where $\alpha$ is a
real constant, Eq.~(\ref{gp}) takes the form
\begin{equation}
\Psi_{xx}+|\Psi |^2
{\Psi}+\left({\alpha^2\over4}\,x^2-{\alpha\over2}\,i+c_3(t)+c_4(t)\,x\right)\,
\Psi  +i\, \Psi_{t}=0 \label{gps2}.
\end{equation}
With $c_3(t)=c_4(t)=0$ this will be the typical GPE with an expulsive harmonic
potential
and damping.\\\\
{\bf Special case III: Harmonic potential and time-dependent
nonlinearity}\\ For $g(x,t)=e^{\alpha\,t}$,
$c_1(t)=c_2(t)=e^{2\alpha\,t}$, Eq.~(\ref{gp}) takes the form
\begin{equation}
\Psi_{xx}+e^{\alpha\,t}\,|\Psi |^2
{\Psi}+\left({\alpha^2\over4}\,x^2+c_3(t)+c_4(t)\,x\right)\, \Psi
 +i\, \Psi_{t}=0 \label{gps3}.
\end{equation}
With $c_3(t)=c_4(t)=0$ this will be the GPE with harmonic potential and
nonlinearity growing exponentially with time. This equation was shown to be
integrable and exact solitonic solution were obtained by
Ref.~\cite{liang}.\\\\
{\bf Special case IV: $x^n$-dependent coefficients}\\
For $g(x,t)=x^n$, where $n$ is an integer, and $c_1(t)=c_2(t)=1$, Eq.~(\ref{gp}) takes the form\\
\begin{equation}
x^{-2n}\,\Psi_{xx}+x^n\,|\Psi |^2
{\Psi}+\left(-{1\over4}n(n+2)x^{-2(n+1)}+{c_3(t)\over
n+1}\,x^{n+1}+c_4(t)\,x\right)\, \Psi  +i\, \Psi_{t}=0 \label{gps4}.
\end{equation}
This is our first counter example to the conclusion of the
Painlev$\rm\acute{e}$ test; an integrable NLS equation with space-dependent coefficients. \\\\
{\bf Special case V: Time-dependent coefficients}\\
For the
special case of $g(x,t)=g(t)$, Eqs.~(\ref{fg}) and (\ref{gam})
result in $f(x,t)=f(t)$ and $\gamma(x,t)=\gamma(t)$, where $f(t)$
and $\gamma(t)$ are now independent. In this case, Eq.~(\ref{vx5})
leads to
\begin{eqnarray}
v(x,t)&=&\frac{-g^2 {\dot f}^2+f g \left(g {\ddot f}+{\dot f}
\left(2 g \gamma -{\dot g}\right)\right)+f^2 \left(2 {\dot g}^2-g
   \left({\ddot g}+4 \gamma  {\dot g}\right)+2 g^2 \left({\dot\gamma}+2 \gamma ^2\right)\right)}{4 f^3
   g^2}\,x^2\nonumber\\&+&c_3(t)\,x+c_4(t)
   \label{vx3},
\end{eqnarray}
where $c_{3}(t)$ and $c_4(t)$ are arbitrary. Comparing this
expression with $v(x,t)=v_2(t)\,x^2+v_1(t)\,x+v_3$, we get
\begin{eqnarray}
&-&f g^2 {\ddot f}+f g {\dot f} {\dot g}-2 f g^2 \gamma  {\dot
f}+g^2 {\dot f}^2+f^2 g {\ddot g}+4 f^2 g \gamma
   {\dot g}-2 f^2 {\dot g}^2\nonumber\\&+&4 {v_2} f^3 g^2-2 f^2 g^2 {\dot\gamma}-4 f^2 g^2 \gamma
   ^2=0.
\end{eqnarray}
This is the integrability condition (22) obtained by He {\it et al.}
\cite{he}. For the special case of $\gamma(t)=0$, we get
\begin{equation}
4 f^3 g^2\,v_2+{ f g ({\dot f} {\dot g}+f {\ddot g})+g^2 ({\dot
f}^2-f {\ddot f})-2 f^2 {\dot g}^2}=0\label{compt2},
\end{equation}
which is the integrability condition (26) of He {\it et al.}


\section{Similarity Transformation method}
\label{simi_sec}

This is a method that is used frequently in the literature to generate new
solutions of a nonautonomous NLS equation from those of the standard
homogeneous NLS equation \cite{sim1,sim2,sim3,raman}.  By transforming the
coordinates and the wave function such that the nonautonomous NLS equation is
transformed into the standard NLS equation, one can use the same transformation
to obtain solutions of the nonautonomous NLS from the solutions of the standard
NLS. Here, we exploit this method to investigate the integrability of NLS
equations. It turns out that for the transformation to be possible, the
coefficients of the nonautonomous NLS must satisfy certain integrability
conditions.

The integrability condition (\ref{compt}) can be obtained by requiring the
transformation
\begin{equation}
\Psi(x,t)=\exp{\left({\over}\beta(x,t)+i\,\theta(x,t)\right)}\,Q(X(x,t))
\label{trans},
\end{equation}
where $\beta(x,t)$, $\theta(x,t)$, and $X(x,t)$ being real functions, to
transform Eq.~(\ref{gp}) into the following time-independent homogeneous
equation
\begin{equation}
p(x,t)\left({\over}\epsilon\,Q_{XX}(X)+\delta\,|Q(X)|^2\,Q(X)\right)=0
\label{homogpe},
\end{equation}
where $p(x,t)$ is in general a complex function that will be determined in
terms of $\beta(x,t)$, $\theta(x,t)$, and $X(x,t)$, and $\epsilon$ and $\delta$
are real constants. Substituting this form of $\Psi(x,t)$ in Eq.~(\ref{gp}), it
takes the form
\begin{eqnarray}
&{}&e^{2 \beta }\, g\, Q(X) |Q(X)|^2+f\, X_x^2\, Q''(X)+ \left[i
X_{t}+f \,\left(2
   X_x \left(\beta _x+i \theta _x\right)+X_{xx}\right)\right]\,Q'(X)\nonumber\\
   &+& \left[v+i\,\gamma+i\, \beta
   _{t}-\theta _{t}+f \left(\left(\beta _x+i \theta _x\right)^2+\beta _{xx}+i
   \theta _{xx}\right)\right]\,Q(X)=0
   \label{homogpe2}.
\end{eqnarray}
Comparing this equation with Eq.~(\ref{homogpe}), we get 8 coupled differential
equations for the unknown functions. These equations can be solved resulting in
rather lengthy expressions for the functions $f(x,t)$, $\theta(x,t)$, and
$v(x,t)$ in terms of $\beta(x,t)$ and arbitrary functions of $t$, as presented
in detail in Appendix~\ref{appendix_b}. We also show in
Appendix~\ref{appendix_b} that in the special case of $\beta(x,t)=\beta(t)$ and
$\gamma(x,t)=0$, the expression for $v(x,t)$ simplifies to
\begin{eqnarray}
v(x,t)&=&{{\dot c}_5 }+\frac{{c_6}^2 \delta ^2 {f} {{\dot c}_2}^2}{4
\epsilon ^2 g^2}\nonumber\\&-&\frac{{c_6} \delta \left({f} g
   {\ddot c}_2+{{\dot c}_2} \left(g {{\dot f}}-2 {f} {\dot g}
   \right)\right)}{2 \epsilon  {f} g^2}\,x\nonumber\\&+&\frac{
   -{f} g \left({\dot f} {\dot g}+{f} {\ddot g}\right)+g^2 \left({f}
   {\ddot f}-{\dot f}^2\right)+2 {f}^2 {\dot g}^2}{4 {f}^3
   g^2}\,x^2,
\end{eqnarray}
where $c_6$ is an arbitrary real constant, and $c_2(t)$ and $c_5(t)$ are
arbitrary real functions. Equating the coefficient of the $x^2$-term of the
previous equation with $v_2$, we retrieve the integrability condition
Eq.~(\ref{compt}) of the Painlev$\rm\acute{e}$ analysis and the Lax Pair
method.

As a second counter example on the integrability of Eq.~(\ref{gp}) with time-
and space-dependent coefficients, we take the special case of
$\beta(x,t)=-(1/2)\log{x}$ and $\gamma(x,t)=0$. Substituting this in $f(x,t)$,
$g(x,t)$, and $v(x,t)$, Eq.~(\ref{gp}) takes the form
\begin{equation}
i\, \Psi_t+\Psi_{xx}+  t^2\, x^3\, |\Psi |^2\,\Psi+\left({3x^2\over
16t^2}-{3\over4x^2}\right)\,\Psi=0\label{gps10},
\end{equation}
where we have set, for simplicity, $\delta=\epsilon=1$. An exact
solution of the homogeneous NLS, Eq.~(\ref{homogpe}), is $Q(X)=-{\rm
sn}(X/\sqrt{2}\,\,|\,-1)$. Applying the Similarity Transformation,
we get a solution of Eq.~(\ref{gps10}), namely
$\Psi(x,t)=\exp{(-i\,x^2/8\,t)}\,{\rm
sn}(t\,x^2/\sqrt{8}\,|-1)/\sqrt{x}$, where ${\rm sn}(x\,|\,m)$ is
the Jacobian elliptic function of modulus $m$.

\section{Integrability conditions of two- and three-dimensional NLS equations}
\label{2d-3d} For the one-dimensional NLS equation, Eq.~(\ref{gp}),
to be integrable, the nonlinearity coefficient, $g(x,t)$, and the
dispersion coefficient, $f(x,t)$, should be related according to the
integrability condition Eq.~(\ref{fg}). It is expected that an
additional term proportional to $\Psi_x(x,t)$ in the NLS equation
would affect this relation between $g(x,t)$ and $f(x,t)$, such that
a wider class of integrable NLS would be obtained. In other words,
it is hoped that one-dimensional NLS equations which are
nonintegrable to become integrable as a result of this addition. In
addition, with this term the NLS corresponds to two- and
three-dimensional physical systems. Furthermore, such an addition
would allow for the consideration of position-dependent effective
mass situations.

We choose the Lax Pair method to investigate the integrability of
the following general NLS equation
\begin{equation}
f(x,t)\, \Psi_{xx}(x,t)+h(x,t)\,\Psi_x(x,t)+g(x,t)\, |\Psi (x,t)|^2
{\Psi}(x,t)+v(x,t)\, \Psi (x,t)+i \,\gamma (x,t)\, \Psi (x,t) +i\,
\Psi_{t}(x,t)=0 \label{gp23d},
\end{equation}
where $h(x,t)$ is a real function.

Employing the Lax Pair method in a similar manner as explained in
section~\ref{lax_sec} and Appendix~\ref{appendix_a}, we obtain the
following integrability conditions
\begin{equation}
f(x,t)={c_1(t)\,H(x,t)\over g(x,t)^2}\label{fg32d},
\end{equation}
\begin{equation}
\gamma(x,t)=\frac{g_t(x,t)}{g(x,t)}-{1\over2} \frac{ {\dot c_2(t)}}{
c_2(t)}-{1\over4}\frac{H_t(x,t)}{H(x,t)} \label{gam32d},
\end{equation}
and
\begin{eqnarray}
&-&4 f g^3 H^2 \left(f_t \left(2 g \gamma
   -g_{t}\right)+f_{tt} g\right)+4 f_t^2 g^4
   H^2+4 f^3 g^3 H \left(v_{x} \left(g H_{x}-2
   g_{x} H\right)+2 g H v_{xx}\right)\nonumber\\
   &+&f^4
   \left[{\over}-g^2H_{x} \left(3 \left(g g_{xx}-2
   g_{x}^2\right) H_{x}+g g_{x} H_{xx}\right)-2
   g H \left(48 g_{x}^3 H_{x}-10 g g_{x}^2
   H_{xx}+g g_{x} \left(g H_{xxx}-42 g_{xx}
   H_{x}\right)\right.\right.\nonumber\\
   &+&\left.\left.2 g^2 \left(2 g_{xx} H_{xx}+3
   g_{xxx} H_{x}\right)\right)+4 \left(36 g_{x}^4-48 g
   g_{xx} g_{x}^2+10 g^2 g_{xxx} g_{x}+g^2
   \left(6 g_{xx}^2-g g^{(4)}\right)\right) H^2{\over}\right]\nonumber\\
   &+&4
   f^2 g^2 H^2 \left[g \left(4 g_{t} \gamma
   +g_{tt}\right)-2 g_{t}^2-2 g^2 \left(\gamma _{t}+2
   \gamma ^2\right)\right)]=0,
   \label{vx523d}
\end{eqnarray}
where
\begin{equation}
H(x,t)=\exp{\left(\int{2h(x,t)\over f(x,t)}\,dx\right)}.
\end{equation}
As a first check, we confirm that the above integrability conditions
indeed reduce to those of the previous sections,
Eqs.~(\ref{fg}-\ref{vx5}) for $H(x,t)=1$ or equivalently $h(x,t)=0$.
These integrability conditions give all coefficients of the NLS
equation in terms of two arbitrary functions $g(x,t)$ and $h(x,t)$.
In the following, we present some interesting special cases.\\\\
{\bf Special case I: Constant effective mass}\\
For $g^2(x,t)=H(x,t)=x^n$, we get the following integrable NLS
equation
\begin{equation}
i\, \Psi_{t}(x,t)+\Psi_{xx}(x,t)+{n\over x}\,\Psi_x(x,t)+x^n\, |\Psi
(x,t)|^2
{\Psi}(x,t)+\left(c_3(t)+c_4(t)\,x+{n(n-2)\over4x^2}\right)\, \Psi
(x,t) =0 \label{gp23ds1},
\end{equation}
where, $c_3(t)$ and $c_4(t)$ are arbitrary real functions. This
equation corresponds to a NLS in $n+1$ spacial dimensions with
constant effective mass. Of particular importance, are the cases of
$n=1$ and $n=2$, corresponding to two and three dimensions,
respectively, with power law for the strength of the interatomic
interaction, $x^n$, and an effective potential that includes the
centripetal part with a suitably-defined angular momentum.\\\\
{\bf Special case II: Power-law effective mass}\\
For $g^2(x,t)=x^p$ and $H(x,t)=x^q$, the integrable NLS takes the
form
\begin{eqnarray}
&&i\,\Psi_{t}(x,t)+x^{q-2p}\Psi_{xx}(x,t)+{1\over
2}\,q\,x^{q-2p-1}\,\Psi_x(x,t)+x^p\, |\Psi (x,t)|^2
{\Psi}(x,t)\nonumber\\
&+&\left(c_3(t)+{2c_4(t)\over2+2p-q}\,x^{1+p-q/2}-{1\over4}\,p(2+p-q)\,x^{q-2p-2}\right)\,
\Psi (x,t) =0 \label{gp23ds2}.
\end{eqnarray}
The position-dependent effective mass appears in the NLS equation
through the second derivative operator, which in this case will be
given explicitly by
\begin{equation}
-{d\over dx}\left({1\over2m(x)}{d\Psi(x,t)\over
dx}\right)=x^{q-2p}\Psi_{xx}(x,t)+{1\over
2}\,q\,x^{q-2p-1}\,\Psi_x(x,t).
\end{equation}
This can be satisfied for $q=4p$, which corresponds to an effective
mass $m(x)=-x^{-2p}/2$, and Eq.~(\ref{gp23ds2}) simplifies to
\begin{eqnarray}
&&i\,\Psi_{t}(x,t)+x^{2p}\Psi_{xx}(x,t)+2p\,x^{2p-1}\,\Psi_x(x,t)+x^p\,
|\Psi (x,t)|^2
{\Psi}(x,t)\nonumber\\
&+&\left(c_3(t)+{c_4(t)\over1-2p}\,x^{1-p}-{1\over4}\,p\,(2-3p)\,x^{2(p-1)}\right)\,
\Psi (x,t) =0 \label{gp23ds2}.
\end{eqnarray}

\section{Conclusions and Outlook}
\label{conc_sec}

Using the Lax Pair and Similarity Transformation methods, we have
shown that, in contrast to the results of the Painlev$\rm\acute{e}$
analysis, Eq.~(\ref{gp}) is integrable for time- and space-dependent
coefficients. This is of course not surprising since it is
well-known that an equation can be integrable in one sense and
nonintegrable in another sense \cite{xxx}. For the special case when
the coefficients of Eq.~(\ref{gp}) are only time-dependent, the
integrabilty conditions of the three methods become identical.

The general integrability conditions found here, Eqs.~(\ref{vx5})
and (\ref{vx523d}), would be useful for cases of space-dependent
dispersion such as optical solitons propagating in a medium of
space-dependent refractive index.

It is in the nature of the Lax Pair method that new solutions are obtained
using old (seed) solutions. Therefore, the integrability conditions imposed by
the Lax Pair method are conditions on the possibility of mapping solutions of
the nonautonomous NLS equation into other solutions of the same NLS. On the
other hand, the integrability conditions imposed by the Similarity
Transformation method are conditions on the possibility of mapping solutions of
the homogeneous NLS equation into those of the nonautonomous NLS. The
integrability conditions of the Painlev$\rm\acute{e}$ method have even more
fundamentally different origin related to the {\it movable} singularities of
the NLS equation. The fact that these three fundamentally different methods
generate the same integrability condition for the case of time-dependent
coefficients of the NLS equation, suggests two illuminating ideas which require
further investigation. First, the three seemingly fundamentally different
methods may be after all not so different such that they can be unified in a
single transformation method. Secondly, it seems that - at least for the case
of time-dependent coefficients - integrability is an intrinsic property of the
NLS equation, which is independent of the method used to reveal it.

Finally, it is worth mentioning that all of the complicated
calculations, including the Lax Pair search, simplifying, and
solving the compatibility equations, that led to the results of this
paper were facilitated using symbolic programming with the computer
program Mathematica 7~\cite{math}.

\appendix
\section{Integrability conditions of the Lax Pair method}
\label{appendix_a} The Lax Pair {\bf U} and {\bf V} are expanded in powers of
$\Psi(x,t)$ and its derivatives, as follows\\\\
\begin{math}
{\bf U}=\left(
\begin{array}{cc}
 {u_{11}} & {u_{12}} \\
 {u_{21}} & {u_{22}}
\end{array}
\right),\hspace{2cm} {\bf V}=\left(
\begin{array}{cc}
 {v_{11}} & {v_{12}} \\
 {v_{21}} & {v_{22}}
\end{array}
\right),\\\\{\rm where}\\
   {u_{11}}={f_1}(x,t)+{f_2}(x,t)\, \Psi (x,t),\\
   {u_{12}}={f_3}(x,t)+{f_4}(x,t)\, \Psi (x,t),\\
   {u_{21}}={f_5}(x,t)+{f_6}(x,t) \Psi^*(x,t),\\
   {u_{22}}={f_7}(x,t)+{f_8}(x,t) \Psi ^*(x,t),\\
   {v_{11}}={g_1}(x,t)+{g_2}(x,t)\,
   \Psi (x,t)+{g_3}(x,t)\, \Psi _x(x,t)+{g_4}(x,t) \Psi (x,t)\, \Psi ^*(x,t),\\
   {v_{12}}={g_5}(x,t)+{g_6}(x,t)\,
   \Psi (x,t)+{g_7}(x,t)\, \Psi_x(x,t)+{g_8}(x,t) \Psi (x,t)\, \Psi ^*(x,t),\\
   {v_{21}}={g_9}(x,t)+{g_{10}}(x,t) \Psi ^*(x,t)+{g_{11}}(x,t)\,
   \Psi _x^*(x,t)+{g_{12}}(x,t)\, \Psi (x,t) \Psi^*(x,t),\\
   {v_{22}}={g_{13}}(x,t)+{g_{14}}(x,t) \Psi ^*(x,t)+{g_{15}}(x,t)\,
   \Psi _x^*(x,t)+{g_{16}}(x,t) \Psi (x,t)\, \Psi
   ^*(x,t),
\end{math}\\
and $f_{1-8}(x,t)$ and $g_{1-16}(x,t)$ are unknown functions.

The compatibility condition
\begin{equation}
{\bf U}_t-{\bf V}_x+[{\bf U},{\bf V}]={\bf0}=\left(
\begin{array}{cc}
 0& p_1(x,t)\,F[\Psi] \\
 p_2(x,t)\,F^*[\Psi^*]&0
\end{array}
\right),
\end{equation}\\\\
with $p_1(x,t)$ and $p_2(x,t)$ being unknown functions, requires
\begin{math}\\\\
{f_2}={g_3}={f_8}={g_{15}}={g_8}={g_{12}}={g_{14}}={g_2}={f_3}={f_5}={
   g_9}={g_5}=0, {f_4}=i\,
   {p_1}, {f_6}=-i\,
   {p_2}, {g_7}=-f\,
{p_1}, {g_{11}}=-f\,
   {p_2}, {g_4}=-g_{16}=-i f\, {p_1}\, {p_2},
   \end{math}\\\\
   and\\\\
\begin{equation}
{f_1}_t-{g_1}_x\label{eq2}=0,
\end{equation}

\begin{equation}
{f_7}_t-{g_{13}}_x\label{eq8}=0,
\end{equation}

\begin{equation}
2 f\, {p_1}\, {p_2}+g\label{eq5}=0,
\end{equation}

\begin{equation}
f_x\, {p_1}-f\, {p_1} ({f_1}-{f_7})+f\, {p_1}_x-{g_6}\label{eq3}=0,
\end{equation}

\begin{equation}
f_x {p_2}+f\, {p_2} ({f_1}-{f_7})+f\, {p_2}_x-{g_{10}}\label{eq7}=0,
\end{equation}

\begin{equation}
{g_6} ({f_1}-{f_7})-i {p_1} ({g_1}-{g_{13}}-i v+\gamma )-{g_6}_x+i
   {p_1}_t\label{eq4}=0,
\end{equation}

\begin{equation}
{g_{10}} ({f_1}-{f_7})+i {p_2} ({g_1}-{g_{13}}-i v-\gamma )+{g_{10}}_x+i
   {p_2}_t\label{eq6}=0,
\end{equation}

\begin{equation}
(f\, {p_1}\, {p_2})_x+{g_{10}}\, {p_1}+{g_6} \,{p_2}\label{eq1}=0.
\end{equation}

\subsection{Deriving a relation between $f(x,t)$ and $g(x,t)$}
Substituting for $f\,p_1\,p_2$ from Eq.~(\ref{eq5}) in Eq.~(\ref{eq1}), the
latter simplifies to
\begin{equation}
-{1\over2}g_x+g_{10}\,p_1+g_6\,p_2=0\label{eq1000}.
\end{equation}
Multiplying Eq.~(\ref{eq3}) by $p_2$ and Eq.~(\ref{eq7}) by $p_1$, and then
adding, we get
\begin{eqnarray}
&{}&p_2\, (f\, {p_1})_x+{p_1}(f\,{p_2})_x-p_2\,g_6-p_1\,g_{10}\nonumber\\
&=&(p_1\, p_2\,f)_x+{p_1}\,p_2\,f_x-p_2\,g_6-p_1\,g_{10}=0.
\end{eqnarray}
Substituting for $p_1\,p_2$, $(f\,p_1\,p_2)_x$, and $p_2\,g_6+p_1\,g_{10}$ from
Eqs.~(\ref{eq5}), (\ref{eq1}) and (\ref{eq1000}), respectively, the last
equation takes the form
\begin{equation}
{f_x\over f}=-2\,{g_x\over g},
\end{equation}
which results in
\begin{equation}
f(x,t)={c(t)\over g(x,t)^2}\label{fg1},
\end{equation}
where $c(t)$ is arbitrary.

\subsection{Deriving a relation between $v(x,t)$ and $g(x,t)$}
Multiplying Eq.~(\ref{eq3}) by $p_2$ and Eq.~(\ref{eq7}) by $p_1$, and then
subtracting, we get
\begin{equation}
f_1-f_7=-{g_6\over f\,p_1}+{1\over2}{\partial\over\partial x}\log{p_1\over
g}+{f\,p_1\,{p_2}_x\over g} \label{eq1001}.
\end{equation}
Multiplying Eq.~(\ref{eq4}) by $p_2$ and Eq.~(\ref{eq6}) by $p_1$, and then
subtracting, we get
\begin{equation}
g_6={p_1\over g}\left[
c_1+i\,k_{1i}-{c\,g_x\over2\,g^2}+{i\over2}\int{\left(3\,g_t-{(2c\,\gamma+{\dot
c})\,g\over c}\right)\,dx} \right].
\end{equation}
To obtain the last equation in this form, we have substituted for $f_1$,
$g_{10}$, $p_2$, and $f$ from Eqs.~(\ref{eq1001}), (\ref{eq1000}), (\ref{eq5}),
and (\ref{fg1}), respectively. Here $c_1(t)$, $k_{1r}(t)$, and $k_{1i}(t)$ are
arbitrary real functions that arise from two integrations over $x$. The last
two being the real and imaginary parts of a complex function. The function
$c_1(t)+k_{1r}(t)$ appearing in the last equation represents another arbitrary
constant. Therefore, we can set, without loss of generality, $k_{1r}(t)=0$.

Multiplying Eq.~(\ref{eq4}) by $p_2$ and Eq.~(\ref{eq6}) by $p_1$, and then
adding, we get
\begin{equation}
g_1-g_{13}=i\,v+{i\,g^2\,g_6^2\over c\,p_1^2}+{{\dot
c}\over2\,c}+{2\,g\,{p_1}_t+2\,i\,g_6\,g_x-3\,p_1\,g_t\over2\,g\,p_1}-{i\,c\,(g\,g_{xx}-2\,g_x^2)\over2\,g^4}
\label{eq1002}.
\end{equation}
To obtain the last equation in this form, we have substituted for $f_1$,
$g_{10}$, $p_2$, and $f$ from Eqs.~(\ref{eq1001}), (\ref{eq1000}), (\ref{eq5}),
and (\ref{fg1}), respectively. Subtracting Eq.~(\ref{eq8}) from
Eq.~(\ref{eq2}), and substituting for $f_1-f_7$ and $g_1-g_{13}$ from
Eqs.~(\ref{eq1001}) and (\ref{eq1002}), we get
\begin{eqnarray}
v_{x}(x,t)&=&-\frac{c} {2g}\frac{\partial
   ^3}{\partial x^3}\,{1\over g}+{2 { k}_{1i}
g_{t}\over c}-{g\over c}
   (2 { k}_{1i} \gamma +{\dot k}_{1i})\nonumber\\
   &+&\frac{g_{t}-g \gamma }{c}\int (3
   g_{t}-{{g\over c} (2 c \gamma
   +{\dot c})}) \, dx\nonumber\\&+&\frac{g}{2c^3} \int \left(c
   \left(g_{t} \left(2 c \gamma +{\dot c}\right)-3 c
   g_{tt}\right)+g \left(c \left(2 c \gamma
   _{t}+{\ddot c}\right)-{\dot c}^2\right)\right) \, dx\nonumber\\&+&i\,\left(\frac{
   g}{c} (2 {c_1} \gamma
   +{\dot c}_1)-{2\over c} {c_1}
   g_{t}\right)\label{vx}.
\end{eqnarray}
Since $v$ is assumed to be real, the imaginary part of the last equation must
vanish. This results in the following condition on $\gamma$
\begin{equation}
\left(2 {c_1} \gamma
   +{{\dot c}_1}\right)\,g-2 {c_1}
   g_t=0,
\end{equation}
which gives
\begin{equation}
\gamma(x,t)=\frac{g_t(x,t)}{g(x,t)}-{1\over2} \frac{ {\dot c_1(t)}}{ c_1(t)}
\label{gam1}.
\end{equation}
Substituting this expression for $\gamma(x,t)$ in Eq.~(\ref{vx}), results in
following integrability condition for $v_x(x,t)$
\begin{eqnarray}
v_{x}(x,t)&=&\frac{g}{2 c^3 {c_1}^2} \left[c^2 {c_1} {\dot c}_1 \int
   \left(g
   (\frac{{\dot c}_1}{{c_1}}-\frac{{\dot c}}{c})+g_t\right) \, dx\right.
   \nonumber\\&+&\left.\int \left({c_1}^2 (c ({\dot c}
   g_t-c g_{tt})+g (c
   {\ddot c}-{\dot c}^2))+c^2 {\dot c}_1^2 g-c^2 {c_1}
   ({\ddot c}_1 g+{\dot c}_1 g_t)\right) \,
   dx\right]\nonumber\\&-&\frac{c} {2\,g}\frac{\partial
   ^3}{\partial x^3}\,{1\over g}-\frac{
   g ({c_1} {{\dot k}_{1i}}-{k_{1i}}
   {\dot c}_1)}{{ c\,c_1}}
   \label{vx4}.
\end{eqnarray}

This condition gives $v_x(x,t)$ in terms of $c(t)$, $c_1(t)$ and $g(x,t)$.
Using Eqs.~(\ref{fg1}) and (\ref{gam1}) to substitute for $c(t)$ and $c_1(t)$,
the last equation gives $v_x(x,t)$ in terms of $f(x,t)$, $\gamma(x,t)$,
$g(x,t)$ and the arbitrary function $k_{1i}(t)$
\begin{eqnarray}
v_{x}(x,t)&=&\frac{1}{2 f^3
   g^5}\int g^4 \left(f \left(g_t
   \left(f_t+2 f \gamma \right)-f
   g_{tt}\right)+g \left(f \left(f_{tt}+2 f \gamma
   _t\right)-f_t^2\right)\right) \, dx\nonumber\\
   &+&2 f^2 g^3
   \left(g_t-g \gamma \right) \int(g
   (-\frac{f_t}{f}-2 \gamma
   )+g_t) \, dx\nonumber\\
   &-&{1\over2}f g\frac{\partial
   ^3}{\partial x^3}\,{1\over g}-\frac{2 g (2
   {k_{1i}} \gamma +{{\dot k}_{1i}})-4 {k_{1i}} g_t}{2 f g^2}
   \label{vx2}.
\end{eqnarray}
Combining the two integrals in Eq.~(\ref{vx4}), differentiating with respect to
$x$, and substituting for $c$ and $c_1$ in terms of $f$, $g$ and $\gamma$, the
last condition takes the form
\begin{eqnarray}
&{}&f g^3 \left(f_t \left(g_t-2 g \gamma \right)-f_{tt} g\right)+f_t^2
   g^4+2 f^3 g^3 \left(g v_{xx}-g_{x} v_{x}\right)\nonumber\\&+&f^2 g^2 \left(g \left(4
   g_t \gamma +g_{tt}\right)-2 g_t^2-2 g^2 \left(\gamma _t+2 \gamma
   ^2\right)\right)\nonumber\\&+&f^4 \left(36 g_{x}^4-48 g g_{xx} g_{x}^2+10 g^2 g_{xxx}
   g_{x}+g^2 \left(6 g_{xx}^2-g g^{(4)}\right)\right)=0,
   \label{vx51}
\end{eqnarray}
which is identical to Eq.~(\ref{vx5}).\\

To summarize: We have derived three integrability conditions:\\\\
{\bf 1.} Condition (\ref{fg1}) gives $f(x,t)$ in terms of
$g(x,t)$ and an arbitrary function $c(t)$.\\\\
{\bf 2.} Condition (\ref{gam1}) gives $\gamma(x,t)$ in terms of
$g(x,t)$ and an arbitrary function $c_1(t)$.\\\\
{\bf 3.} Condition (\ref{vx51}) gives $v_x(x,t)$ in terms
of $g(x,t)$, $f(x,t)$ and $\gamma(x,t)$.\\
The three arbitrary functions $c(t)$, $c_1(t)$, and $k_{1i}(t)$ are independent
from each other.

\section{Integrability conditions of the Similarity Transformation method}
\label{appendix_b}

Comparing Eq.~(\ref{homogpe}) with Eq.~(\ref{homogpe2}) we get
\begin{equation}
e^{2 \beta }\, g=\delta\,p\label{eq11},
\end{equation}
\begin{equation}
f\,X_x^2=\epsilon\,p\label{eq12},
\end{equation}
\begin{equation}
 X_{t}+2f \,X_x \theta _x=0\label{eq14},
\end{equation}
\begin{equation}
2X_x\,\beta _x+X_{xx}=0\label{eq15},
\end{equation}
\begin{equation}
\beta _{t}+\gamma+f \left(2\beta _x\,\theta _x+
   \theta _{xx}\right)=0\label{eq16},
\end{equation}
\begin{equation}
v-\theta _{t}+f \left(\beta _x^2- \theta _x^2+\beta _{xx}\right)=0\label{eq17}.
\end{equation}
Solving Eq.~(\ref{eq15}) for $X$, we find
\begin{equation}
X(x,t)=c_1(t)+c_2(t)\,\int{e^{-2\beta(x,t)}\,dx}\label{xeq}.
\end{equation}
Eliminating $p(x,t)$ from Eqs.~(\ref{eq11}) and (\ref{eq12}) and substituting
for $X(x,t)$ from the previous equation, we find a compatability condition that
combines $f(x,t)$ and $g(x,t)$, namely
\begin{equation}
g(x,t)={\delta\over\epsilon}\,c_2(t)^2\,e^{-6\beta(x,t)}\,f(x,t)\label{geq}.
\end{equation}
Solving Eq.~(\ref{eq14}) for $\theta(x,t)$, we get
\begin{eqnarray}
\theta (x,t)=-\int \frac{ e^{2 \beta}\left(\int \,e^{-2 \beta}\left({\dot c}_2
-2 \,{c_2}
 \beta
   _{t}\right) \, dx+{\dot c}_1\right)}{2\,{c_2} f} \, dx+{c_\theta
   }(t)\label{thetaeq},
\end{eqnarray}
where $c_\theta(t)$ is arbitrary. Solving Eq.~(\ref{eq16}) for $f(x,t)$, we get
\begin{equation}
f(x,t)={c_f(t)} \exp \left(\int \frac{e^{-2 \beta } \left(4 e^{2 \beta } \beta
_{x} \left(\int e^{-2 \beta
   } \left({\dot c}_2-2 {c_2} \beta _{t}\right) \, dx+{\dot c}_1\right)-4 {c_2} \beta
   _{t}+{\dot c}_2-2\,c_2\,\gamma\right)}{\int e^{-2 \beta } \left({\dot c}_2-2 {c_2} \beta _{t}\right) \,
   dx+{\dot c}_1} \, dx\right)\label{feq},
\end{equation}
where $c_f(t)$ is arbitrary.

\subsection{Special case I} To obtain the special case of time-dependent coefficients of the NLS equation,
namely $f(x,t)=f(t)$, $g(x,t)=g(t)$ , we assume $\beta(x,t)=\beta(t)$ and
$\gamma(x,t)=\gamma(t)$. With these assumptions, Eq.~(\ref{feq}) takes the form
\begin{equation}
f(x,t)=\left((2 {c_2}\, {\dot b}-{\dot c}_2)\,x-e^{2 b} {\dot
c}_1\right)^{\frac{2 {c_2} \left(2 {\dot b}+\gamma
   \right)-{\dot c}_2}{2 {c_2} {\dot b}-{\dot c}_2}}.
\end{equation}
The function $f(x,t)$ becomes $x$-independent either by setting the exponent in
the previous equation to zero or by setting the coefficient of $x$ to zero. We
notice that the coefficient of
 $x$ is identical with the denominator of the exponent.
 Therefore, we take the first choice.
 Setting the exponent to zero and solving for $c_2$, we get
\begin{equation}
c_2(t)=c_8\,e^{4\,b+2\int{\gamma\,dt}}\label{c2eq}.
\end{equation}
Substituting this expression for $c_2(t)$ in Eqs.~(\ref{feq}) and
(\ref{thetaeq}), we get $f(x,t)=c_f(t)\equiv f(t)$ and
\begin{equation}
\theta (x,t)={c_\theta }-\frac{x \left(x {\dot b}+{{\dot c}_1\,} e^{-2 (b+\int
\gamma  \, dt)}/c_8\right)+x^2 \gamma }{2{f}}.
\end{equation}
Finally, the function $v(x,t)$ can be obtained from Eq.~(\ref{eq17})
\begin{eqnarray}
v(x,t)&=&\frac{ e^{-2 (b+\int \gamma  \, dt)} \left({\dot c}_1 \left(4 {f}
\left({\dot b}+\gamma
   \right)+{\dot f}\right)-{f} {\ddot c}_1\right)}{2 {c_8} {f}^2}\,x\nonumber\\
   &+&\frac{ {\dot f}
   \left({\dot b}+\gamma \right)+{f} \left(-{\ddot b}+4 \gamma  {\dot b}+2
   {\dot b}^2-{\dot\gamma}+2 \gamma ^2\right)}{2
   {f}^2}\,x^2\nonumber\\&+&\frac{{\dot c}_1^2 e^{-4 (b+\int \gamma  \, dt)}}{4 {c_8}^2 {f}}+{{\dot c}_\theta }
\end{eqnarray}
taking the special case $\gamma=0$ and substituting from Eq.~(\ref{geq}) for
$b=(1/2)\log{(\epsilon\,g/c_8^2\,\delta\,f)}$, the last equation gives
\begin{eqnarray}
v(x,t)&=&{{\dot c}_\theta }+\frac{{c_8}^2 \delta ^2 {f} {{\dot c}_1}^2}{4
\epsilon ^2 g^2}\nonumber\\&-&\frac{{c_8} \delta \left({f} g
   {\ddot c}_1+{{\dot c}_1} \left(g {{\dot f}}-2 {f} {\dot g}
   \right)\right)}{2 \epsilon  {f} g^2}\,x\nonumber\\&+&\frac{
   -{f} g \left({\dot f} {\dot g}+{f} {\ddot g}\right)+g^2 \left({f}
   {\ddot f}-{\dot f}^2\right)+2 {f}^2 {\dot g}^2}{4 {f}^3 g^2}\,x^2
\end{eqnarray}
Equating the coefficient of the $x^2$-term with $v_2$, we retrieve the
integrability condition Eq.~(\ref{compt}) of the Painlev$\rm\acute{e}$ analysis
and the Lax Pair method.

\end{document}